\documentclass[]{article}

\PassOptionsToPackage{numbers, compress}{natbib}

\usepackage[preprint]{neurips_2020}
\usepackage[utf8]{inputenc} % allow utf-8 input
\usepackage[T1]{fontenc}    % use 8-bit T1 fonts
\usepackage{hyperref}       % hyperlinks
\usepackage{url}            % simple URL typesetting
\usepackage{booktabs}       % professional-quality tables
\usepackage{amsfonts}       % blackboard math symbols
\usepackage{nicefrac}       % compact symbols for 1/2, etc.
\usepackage{microtype}      % microtypography
\usepackage{amsmath,amssymb}	
\usepackage{mathtools}
\usepackage{lipsum}
\usepackage[dvipsnames]{xcolor}

\definecolor{linkcolor}{rgb}{0.7752941176470588, 0.22078431372549023, 0.2262745098039215}
\hypersetup{colorlinks=true,
linkcolor=linkcolor,
citecolor=linkcolor,
urlcolor=linkcolor,
linktocpage=true,
pdfproducer=medialab,
}

\title{Semi-parametric $\gamma$-ray modeling with \\ Gaussian processes and variational inference}

\author{%
  Siddharth Mishra-Sharma, Kyle Cranmer\\
  New York University \\
  \texttt{\{sm8383,\,kyle.cranmer\}\,@nyu.edu} \\
}

\begin{document}

\maketitle

\begin{abstract}
  Mismodeling the uncertain, diffuse emission of Galactic origin can seriously bias the characterization of astrophysical $\gamma$-ray data, particularly in the region of the Inner Milky Way where such emission can make up $\gtrsim80\%$ of the photon counts observed at $\sim$\,GeV energies. We introduce a novel class of methods that use Gaussian processes and variational inference to build flexible background and signal models for $\gamma$-ray analyses with the goal of enabling a more robust interpretation of the make-up of the $\gamma$-ray sky, particularly focusing on characterizing potential signals of dark matter in the Galactic Center with data from the \emph{Fermi} telescope.
\end{abstract}

\section{Introduction}
\label{sec:intro}

The nature of dark matter remains a major, persisting mystery in particle physics and cosmology today. One of the primary avenues to search for dark matter is through astrophysical indirect detection---looking for visible byproducts of dark matter annihilation or decay in dark matter-rich regions of the sky (see Refs.~\cite{Slatyer:2017sev,Lisanti:2016jxe} for a review). Indeed such a putative signal was identified in the inner regions of the Milky Way---the Galactic Center---over a decade ago using $\gamma$-ray data from the \emph{Fermi}-LAT space telescope~\cite{Goodenough:2009gk}. The origin of this signal, known as the Galactic Center Excess (GCE), has been hotly debated and remains contentious. The spatial and spectral properties of the signal were shown early on to be compatible with expectation from annihilating dark matter~\cite{Hooper:2010mq,Daylan:2014rsa}. More recently, the statistical properties of the signal were shown to prefer an explanation in terms of an unresolved population of $\gamma$-ray point sources (PSs) rather than annihilating dark matter using the 1-point statistics of photon counts~\cite{Lee:2015fea} and a wavelet decompositions of the GCE signal~\cite{Bartels:2015aea}. The spatial morphology of the signal was subsequently also shown to prefer an astrophysical explanation~\cite{Macias:2016nev,Macias:2019omb,Bartels:2017vsx}.

Robustly characterizing the GCE signal is however complicated by the fact that the Galactic Center region contains significant background emission of diffuse Galactic origin sourced by cosmic rays interacting with the gas, dust, and charged particle populations in the Milky Way. While this Galactic background emission can be modeled using cosmic-ray propagation codes such as Galprop~\cite{galprop,Strong:1999sv} and  Dragon~\cite{Evoli:2016xgn}, uncertainties in the properties of 3-D cosmic-ray transport, as well as the underlying distribution of interstellar gas and dust mean that large uncertainties in our knowledge these background components remain.

While the existence of the GCE has been shown to be generally robust to reasonable variation of the modeled diffuse emission~\cite{Daylan:2014rsa,Calore:2014xka,TheFermi-LAT:2015kwa,Linden:2016rcf}, its statistical interpretation as originating from a population of unresolved PSs can be be susceptible to mismodeling of the Galactic diffuse emission, to the extent of potentially mischaracterizing a dark matter signal as arising from a population of PSs~\cite{Lee:2015fea,Leane:2019uhc}. Reliably characterizing the origin of the GCE therefore requires an improved understanding of the Galactic diffuse contribution. Complementary to building better models of this emission are methods that augment existing models by introducing extra spatial and/or spectral degree of freedom that aim to account for uncertainties in our knowledge of diffuse emission of Galactic origin~\cite{Storm:2017arh,Bartels:2017vsx,Buschmann:2020adf,Chang:2018bpt}. In particular, Refs.~\cite{Storm:2017arh,Bartels:2017vsx} used regularized likelihoods to build adaptive templates for the Galactic diffuse emission. More recently, Ref.~\cite{Buschmann:2020adf} showed that some of the practical issues associated with imperfect background modeling can be alleviated by marginalizing over the large-scale structure of the diffuse model in the basis of spherical harmonics. In this work, we present a complementary approach that uses Gaussian processes and recent advances in variation inference to build flexible models of the Galactic diffuse emission with the goal of robustly characterizing the $\gamma$-ray sky.

\section{Model and inference}
\label{sec:model}

\paragraph{Template regression}

Template regression is a standard technique in astrophysics and cosmology where spatially and/or spectrally binned data is described through a set of spatial templates, binned the same way as the data, each representing the contribution of a particular modeled component, either signal or background. For counts data, Poisson template regression is often employed, where the data is assumed to be a Poisson realization of the modeled emission. Focusing on a single spectral bin with $p$ indexing spatial bins, we have the pixel-wise counts data $d^{p}\sim\mathrm{Pois}\left(\mu^{p}(\boldsymbol{\theta})\right)$, where $\boldsymbol{\theta}$ represents the template parameters. Most commonly, the template parameters correspond to overall normalizations $A_i$ of the individual spatial templates $T_{i}^{p}$,  \emph{i.e.} $\mu^{p}(\boldsymbol{\theta}) = \sum_{i}A_i T_{i}^{p}$. This framework is often employed to search for the evidence of a particular signal (\emph{e.g.}, emission of dark matter origin) in counts data while accounting for uncertainties in the  normalizations of background template parameters in either a Bayesian~\cite{Hoof:2018hyn} of frequentist~\cite{Lisanti:2017qoz} setting.

Beyond Poisson template regression, methods based on the 1-point PDF of photon counts can model the contribution of populations of unresolved PSs to the counts data, where each PS is too dim to be resolved individually but the collective emission from the population can be detected statistically~\cite{Malyshev:2011zi,Lee:2014mza}. These methods can be used in a model comparison setting to differentiate between PS-like (`clumpy') and DM-like (`smooth') origins of a signal following a given spatial morphology. While we focus on the simpler case of Poisson regression in this work, extensions to non-Poissonian template fitting are easily admitted by modifying the likelihood and introducing additional parameters characterizing the contribution of unresolved PS populations~\cite{Mishra-Sharma:2016gis}. 

\paragraph{Augmenting the diffuse template with Gaussian Processes} 

Here, the goal is to give additional freedom to certain background templates---in particular, the Galactic diffuse emission template---in order to enable more robust characterization of the other modeled contributions. We propose doing so using Gaussian processes (GPs), which define a prior distribution over the space of functions such that any collection of function values, evaluated at any collection of points, has a multivariate Gaussian distribution. In particular, $\left[f\left(x_{1}\right), \ldots, f\left(x_{n}\right)\right] \sim \mathcal{N}(m, K)$,
where the covariance function $K_{i j}=k\left(x_{i}, x_{j}\right)$ controls the inductive biases of the GP model, \emph{e.g.} its smoothness and periodicity, and the mean function $m$ is often set to zero (see Refs.~\cite{10.5555/1162254,wilson2014covariance} for a review). Here we employ the Matérn kernel, which generalizes the more common exponential quadratic kernel, and fix the smoothness parameter $\nu = 5/2$, which was found to work well in the present context. The lengthscale and variance of the kernel are hyperparameters of the GP. Since we are interested in modeling processes on (a sub-region of) the celestial sphere, we use the great-circle distance (in angular units) as our distance measure.

Treating the other templates as before using an overall normalization factor, following Ref.~\cite{Buschmann:2020adf} we modulate the Galactic diffuse template (denoted with subscript `dif') by a Gaussian process so that the overall model is described by
\begin{equation}
  d^{p} \sim \mathrm{Pois}\left(\sum_{i\neq\mathrm{dif}} A_i T_{i}^{p} + \exp\left(f^{p}\right)A_\mathrm{dif}T_\mathrm{dif}^{p}\right)
  \label{eq:poisson_gp}
\end{equation}
where $f\sim \mathcal{N}\left(m, K\right)$ is the GP component and an exponential link function is used to ensure positivity of the zero-mean GP component. We fix the non-GP multiplicative coefficient of the diffuse template $A_\mathrm{dif}$ to the best-fit value found through a maximum-likelihood fit of the model without the GP component in our region of interest. We note that setting the GP mean to zero does not imply a zero \emph{predictive} mean, and the GP component has ample freedom to model departures from the mean obtained in the initial fit.

Our ultimate goal is latent function inference, and there is not a distinction between training and test data points here as is often the case in applications of GPs to predictive models. In addition to the fact that the marginal/predictive likelihoods are not analytically tractable in this setting with a non-Gaussian likelihood, the datasets in question can be relatively large with $n_\mathrm{pix} \sim\mathcal{O}(10^4)$ data points. We thus make use of sparse variational Gaussian process (SVGP) methods~\cite{quinonero-candela_unifying_2005} implemented through GPyTorch~\cite{gardner2018gpytorch}, further relying on PyTorch~\cite{NEURIPS2019_9015}.

\paragraph{Variational inference}

Our ultimate goal is to characterize the contribution of various templates and models to $\gamma$-ray data in parallel to learning the structure of the Gaussian process that accounts for large-scale uncertainties in the Galactic diffuse template. Even in the realm of sparse GPs, sampling the posterior distribution of various template parameters in conjunction with learning the GP (hyper)parameters is computationally expensive. Variational inference tackles this issue by approximating the posterior over parameters of interest with a simpler parameterized distribution, known as the variational distribution. The approximate density is then fit by maximizing a lower bound on the marginal log-likelihood, the evidence lower bound (ELBO)~\cite{wingate_automated_2013,hoffman2013stochastic}. 

For the Gaussian process component, the posterior of function values $f^u$ over a smaller number $n_u \leq n_\mathrm{pix}$ of inducing points located at $x_u$ is approximated using a multivariate Gaussian distribution $q(f^{u})=\mathcal N(m_u, K_u)$ with learned mean $m_u$ and covariance $K_u$, and the inducing point locations $x_u$ themselves are learned parameters. 
The strategy of Ref.~\cite{pmlr-v38-hensman15}, implemented in GPyTorch~\cite{gardner2018gpytorch}, is employed which uses a GP that jointly models the distribution of function values at the locations of the inducing points and pixel locations, and then marginalizes out the function values at the inducing points giving the desired variational distribution for the GP function values $f^{p}$ at the pixel locations $x_{p}$, \emph{i.e.} $q(f^{p}) = \int  \mathrm{d}f^{u}\, p(f^{p} | f^{u})\,q(f^{u})$.

Besides the variational treatment of the GP component, a parametric form for the variational distribution over the model parameters characterizing the other spatial templates---in this case, the template normalizations---has to be chosen. A common choice is to again describe the joint variational distribution over all the parameters as a multivariate Gaussian with learned mean and covariance. This choice has several drawbacks for the present application---it restricts the form of the template parameter posteriors to (correlated) Gaussian distributions, and is additionally unable to model correlations between the GP component and template parameters. The ability to model such correlations and more complicated posterior distributions may be especially important in applications beyond the Poisson regression case we consider here---one can imagine, \emph{e.g.}, residuals associated to PS populations to be highly correlated with the GP latent function.

To allow for more expressive posterior distributions for the template parameters $A_i$ and model correlations between the posteriors of the GP and template parameters, we use normalizing flows~\cite{pmlr-v37-rezende15} conditioned on summary statistics $\mathbf{s}(f)$ of the GP samples to model the variational distribution of the template parameters $q(A_i|\mathbf{s}(f))$.
This is done following Ref.~\cite{quinonero-candela_unifying_2005}, starting with a unit Gaussian base distribution with diagonal covariance $\mathcal{N}(0, I)$ and applying a series of inverse autoregressive flow (IAF) transformation---normalizing flows containing masked autoregressive neural networks (NNs)~\cite{germain2015made}---to model the template parameter variational distributions. The autoregressive NNs are augmented to take in additional context variables as inputs in order to condition the template posteriors on variables $\mathbf{s}(f)$ summarizing samples from the GP variational distribution (see Ref.~\cite{10.5555/3045390.3045710} for details). In this work, we use a setup with 4 IAF transformations and autoregressive NNs with 3 hidden layers and 10 times the number of nodes per layer as input (template) parameters. The per-pixel-normalized dot product of the (exponentiated) GP sample $\exp\left(f^p\right)$ with each spatial template as well as the mean and square root of the spatial variance of $\exp\left(f^p\right)$ over the pixel locations $x_p$ are used as context variables for conditioning the transformations in order to capture correlations between the GP and template parameters. We note that additional context variables characterizing the GP sample or even the entire GP map (or a downgraded version of it) may be used as inputs to the NNs in order to capture subtler correlations between the template and GP components, as required. The variational model is constructed using Pyro~\cite{bingham2019pyro}, with GPyTorch~\cite{gardner2018gpytorch} used to to model the variational GP components. 

\section{Tests on simulated data}
\label{sec:experiments}

We use simulated \emph{Fermi}-LAT $\gamma$-ray data to validate our pipeline using the example dataset and templates provided with Refs.~\cite{Mishra-Sharma:2016gis,Buschmann:2020adf}, corresponding to 413 weeks of data in the 2--20\,GeV energy range (see Ref.~\cite{Mishra-Sharma:2016gis} for additional information about the dataset). In addition to the $\gamma$-ray counts data, templates corresponding to resolved PSs from the \emph{Fermi} 3FGL catalog~\cite{Acero:2015hja}, isotropically-uniform emission, emission from the \emph{Fermi} bubbles~\cite{Su:2010qj}, emission from dark matter annihilation following a squared Navarro-Frenk-White (NFW)~\cite{Navarro:1995iw,Navarro:1996gj} spatial profile, and Galactic diffuse emission modeled using either the \texttt{p6v11}~\cite{p6v11} model or the more recent Model O~\cite{Buschmann:2020adf} are provided. The maps are spatially binned using HEALPix~~\cite{Gorski:2004by,Zonca2019} with \texttt{nside}=128. A region of interest (ROI) with latitude cut $|b| > 2^\circ$ and radial cut $r < 20^\circ$ is chosen, and resolved PSs from the 3FGL catalog masked at $0.8^\circ$. This corresponds to a typical GCE analysis ROI with $n_\mathrm{pix}=4234$ total pixels. 

By creating simulated data using one Galactic diffuse model and analyzing it with another diffuse model, we can get a sense of how well we are able to recover the `ground truth' diffuse mismodeling introduced in the setup. We create simulated data as a Poisson realization of the sum of templates in the analysis ROI best-fit to the real \emph{Fermi} data, using Model O to model the Galactic diffuse emission. Two separate templates for the Galactic diffuse emission are independently floated for Model O---one correlated with the gas, accounting for a combination of emission due to bremsstrahlung and neutral pion decay, and another modeling inverse Compton emission (see Ref.~\cite{Buschmann:2020adf} for further details).  We then analyze the simulated data with the \texttt{p6v11} diffuse model, using Poisson regression with a GP modulating the single diffuse emission template as in Eq.~\eqref{eq:poisson_gp}. The GP hyperparameters---its lengthscale and variance---are fixed to respective mean values obtained by performing an exact GP regression of the diffuse model used in the fit with a suite of other diffuse models from Ref.~\cite{Ackermann:2014usa} (not including diffuse Model O, which was used to create the simulated data) in the analysis ROI.

\begin{figure}[!t]
  \centering
  \includegraphics[width=0.98\textwidth]{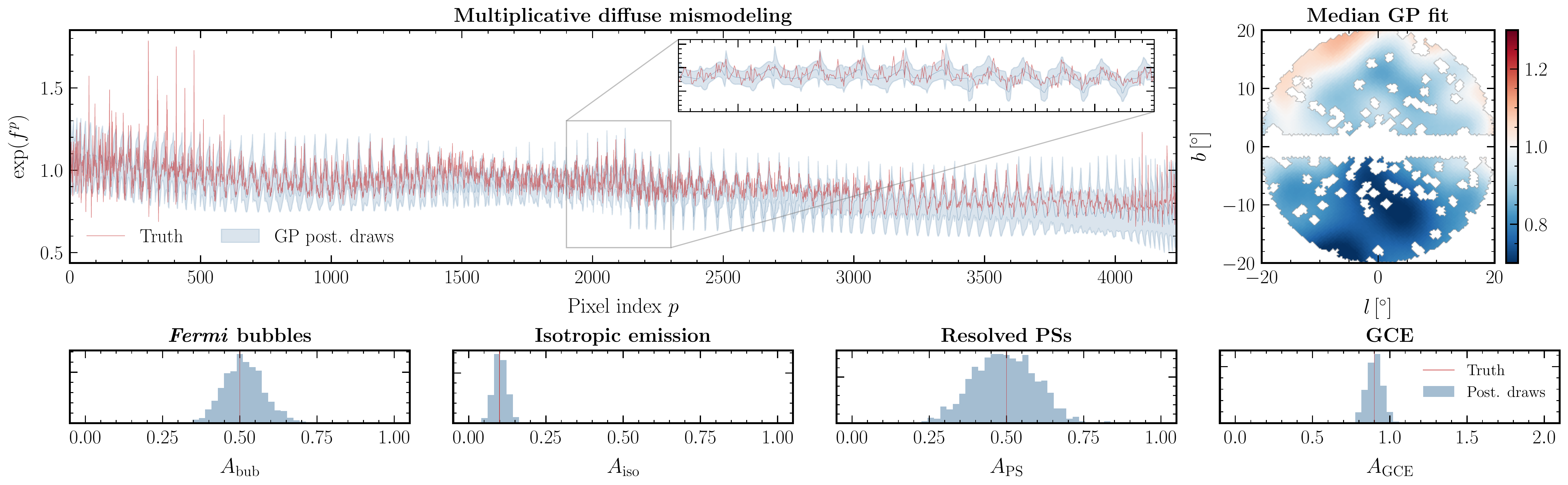}
  \caption{\emph{(Top left)} Pixel-wise 95\% highest-posterior density interval (blue band) and median (blue line) of the GP posterior-predictive distribution, along with the true multiplicative modeling between the Model O (in simulation) and \texttt{p6v11} (in fit) diffuse background models (red line). Pixel indices cross the map left to right, starting from the top. The inset shows a zoomed-in region closest to the Galactic Center $(l, b) = (0^\circ, 0^\circ)$. The Gaussian process is seen to faithfully describe uncertainty in the diffuse model on larger scales. \emph{(Top right)} The median inferred map of multiplicative mismodeling in the analysis region of interest. \emph{(Bottom row)} Samples from the posterior-predictive distributions of the Poissonian template normalizations (blue histograms) and the corresponding ground truths (vertical red lines).}
  \label{fig:experiment}
\end{figure}

Since each pixel is conditionally independent given the model parameters, subsampling can be used to significantly speed up inference while serving as an additional source of stochasticity during training. The model is trained using the Adam optimizer~\cite{DBLP:journals/corr/KingmaB14} with learning rate $\alpha=10^{-3}$ and other parameters set to their default values in the PyTorch implementation, and run for 50,000 iterations with a subsample size of 1500 and 200 inducing points.
Fig.~\ref{fig:experiment} shows the posterior-predictive distributions for the template normalization parameters (bottom row) as well as the pixel-wise median and 95\% highest-posterior density interval of the exponentiated Gaussian process (top left, shown as a function of HEALPix pixel index within the analysis ROI), obtained by taking 1000 samples from the respective predictive distributions. Ground truth values for the template normalizations and multiplicative mismodeling are shown in red. We see that the GP is able to faithfully model the residual large-scale mismodeling in the diffuse emission, and the normalizations of the respective Poissonian templates are correctly recovered. The median inferred GP map is shown in the top right panel.

\section{Conclusions and outlook}
\label{sec:conclusions}

We have introduced a novel method to account for mismodeling in analyses of astrophysical counts data with a particular emphasis on modeling uncertainties in emission of Galactic diffuse origin in $\gamma$-ray analyses. An immediate application of our method is to characterize the statistical nature of the \emph{Fermi} Galactic Center Excess while accounting for large-scale diffuse mismodeling uncertainties, which would require extending our framework to, \emph{e.g.}, use a likelihood based on the the 1-point PDF instead of Eq.~\eqref{eq:poisson_gp}. Beyond their application to Poisson regression and 1-point PDF methods, GPs can also be naturally included as components of machine learning-aided $\gamma$-ray analyses~\cite{List:2020mzd,Caron:2017udl}. Additionally, given uncertainties in the spatial distribution of the GCE signal itself and their potential to bias statistical inference of its nature~\cite{Leane:2020nmi,Leane:2020pfc}, GP-based analyses can be extended to infer the signal morphology in a data-driven manner. 

\section*{Broader Impact}
\label{sec:impact}

Accounting for epistemic uncertainty is crucial for making robust conclusions from data in machine learning applications. This work is part of the broader scientific effort to design and implement techniques that attempt to incorporate deficiencies in our ability to model consequential aspects of real-world data in a principled manner.

We acknowledge the importance of considering the ethical implications of scientific research in general, and machine learning research in particular, as well as of placing both the process and output of scientific research in a broader societal context. We do not believe the present work presents any issues in this regard. 

\begin{ack}
 We thank Lukas Heinrich for collaboration at the early stages of this work. KC is partially supported by NSF grant PHY-1505463m, NSF awards ACI-1450310, OAC-1836650, and OAC-1841471, and the Moore-Sloan Data Science Environment at NYU. SM is supported by the NSF CAREER grant PHY-1554858, NSF grants PHY-1620727 and PHY-1915409, and the Simons Foundation. This work was also supported through the NYU IT High Performance Computing resources, services, and staff expertise. 
We thank the \emph{Fermi} collaboration for making publicly available the $\gamma$-ray data used in this work. This research has made use of NASA's Astrophysics Data System. This research made use of the Astropy~\cite{Robitaille:2013mpa,Price-Whelan:2018hus},
GPyTorch~\cite{gardner2018gpytorch},
HEALPix~\cite{Gorski:2004by,Zonca2019},
IPython~\cite{PER-GRA:2007},
Jupyter~\cite{Kluyver2016JupyterN},
Matplotlib~\cite{Hunter:2007},
NumPy~\cite{harris_array_2020},
Pyro~\cite{bingham2019pyro},
PyTorch~\cite{NEURIPS2019_9015},
SciPy~\cite{2020SciPy-NMeth}, and
Seaborn~\cite{michael_waskom_2017_883859}
software packages.
\end{ack}

\bibliographystyle{apsrev4-1-mod}
\bibliography{fermi-gp}

\end{document}